\documentclass{tgfinal}

\usepackage{amssymb}
\usepackage{graphicx,natbib}

\newcommand{\be}{\begin{equation}}
\newcommand{\ba}{\begin{eqnarray}}
\newcommand{\ee}{\end{equation}}
\newcommand{\ea}{\end{eqnarray}}

\def\lesssim{\mathrel{\hbox{\rlap{\hbox{\lower4pt\hbox{$\sim$}}}\hbox{$<$}}}}
\def\gtrsim{\mathrel{\hbox{\rlap{\hbox{\lower4pt\hbox{$\sim$}}}\hbox{$>$}}}}

\def\simless{\mathbin{\lower 3pt\hbox
   {$\rlap{\raise 5pt\hbox{$\char'074$}}\mathchar''7218$}}}   
\def\simgreat{\mathbin{\lower 3pt\hbox
   {$\rlap{\raise 5pt\hbox{$\char'076$}}\mathchar''7218$}}}   
\def\apj{ApJ}

\def\aap{A\&A}

\def\mnras{MNRAS}

\newcommand{\url}{\tt}%
\begin{document}

\title{Simulating Cosmic Reionization}


\author{
Ilian T. Iliev (University of Z\"urich), 
Paul R. Shapiro (University of Texas at Austin), 
\\Garrelt Mellema (University of Stockholm), 
Hugh Merz (University of Waterloo), 
\\Ue-Li Pen (Canadian Institute for Theoretical Astrophysics, University
of Toronto)} 


\maketitle

 \begin{abstract}
   The first billion years of the history of the Universe, commonly
   referred to as the Cosmic Dark Ages and the Epoch of Reionization,
   constitute a crucial missing link in our understanding of the
   evolution of the intergalactic medium and the formation and
   evolution of galaxies.  As gravity gradually amplified the small
   density contrasts in this dark-matter-dominated universe, structure
   arose hierarchically, with small objects condensing out first, then
   merging with each other to make ever larger objects in a clustered
   pattern in space known as the Cosmic Web.  Less than a hundred
   million years after the Big Bang, the stars forming in the first
   dwarf galaxies began to release ionizing radiation, which leaked out
   of these galaxies and caused ionization fronts to sweep outward
   through the surrounding primordial atomic gas, gradually transforming
   the cold, neutral intergalactic medium into a hot, ionized one.
   This epoch of reionization ended before the universe was a billion
   years old, at a time from which the light now reaching us has its
   wavelength redshifted by about a factor of 7.  Due to the complex
   nature of this global process it is best studied through large-scale
   numerical simulations. A number of large observational efforts
   trying to detect this process directly are currently under way, and
   their success depends critically upon correctly modelling its
   observable signatures.

 Such simulations present considerable challenges, however, related to
 the large dynamic range required and the necessity to perform fast and
 accurate radiative transfer calculations. The hierarchical nature of
 cosmological structure formation in the Cold Dark Matter paradigm
 means that, at these early times, the dominant contributors of ionizing 
 radiation were dwarf galaxies of around one billion Solar masses. These 
 tiny galaxies must be resolved in very large cosmological volumes in 
 order to derive their clustering properties and the corresponding
 observational signatures correctly, which makes this one of the most 
 challenging problems of numerical cosmology.

 We have recently performed the largest and most detailed simulations
 of the formation of early cosmological large-scale structures and
 their radiative feedback leading to cosmic reionization. This was
 achieved by running extremely large, $1728^3$- to $3072^3$-particle
 (5.2 to 29 billion) N-body simulations of the formation of
 the Cosmic Web, with enough particles and
 sufficient force resolution to resolve all the galactic halos with
 total masses larger than 100 million Solar masses in computational
 volumes of $(64/h\,\rm Mpc)^3$ to $(114/h\,\rm Mpc)^3$, respectively. 
 These results
 were then post-processed by propagating the ionizing radiation from all
 (up to millions of) sources by using fast and accurate ray-tracing
 radiative transfer method. For these simulations, we utilized a P$^3$M
 N-body code called CubeP$^3$M and a radiative transfer code called
 C$^2$-Ray, both developed by us. Both codes are parallelized
 using a combination of MPI and OpenMP and to this date have been run
 efficiently on up to 2048 cores (CubeP$^3$M) and up to 10000 cores
 (C$^2$-Ray) on the newly-deployed Sun Constellation Linux Cluster
(Ranger) at the Texas Advanced Computing Center.

 In this paper we briefly present our codes, methods and
 parallelization strategies, discuss our recent simulations and  results
 and outline our future plans.
\end{abstract}
\section{Introduction} 

\begin{figure*}
  \begin{center}
    \includegraphics[width=3.0in]{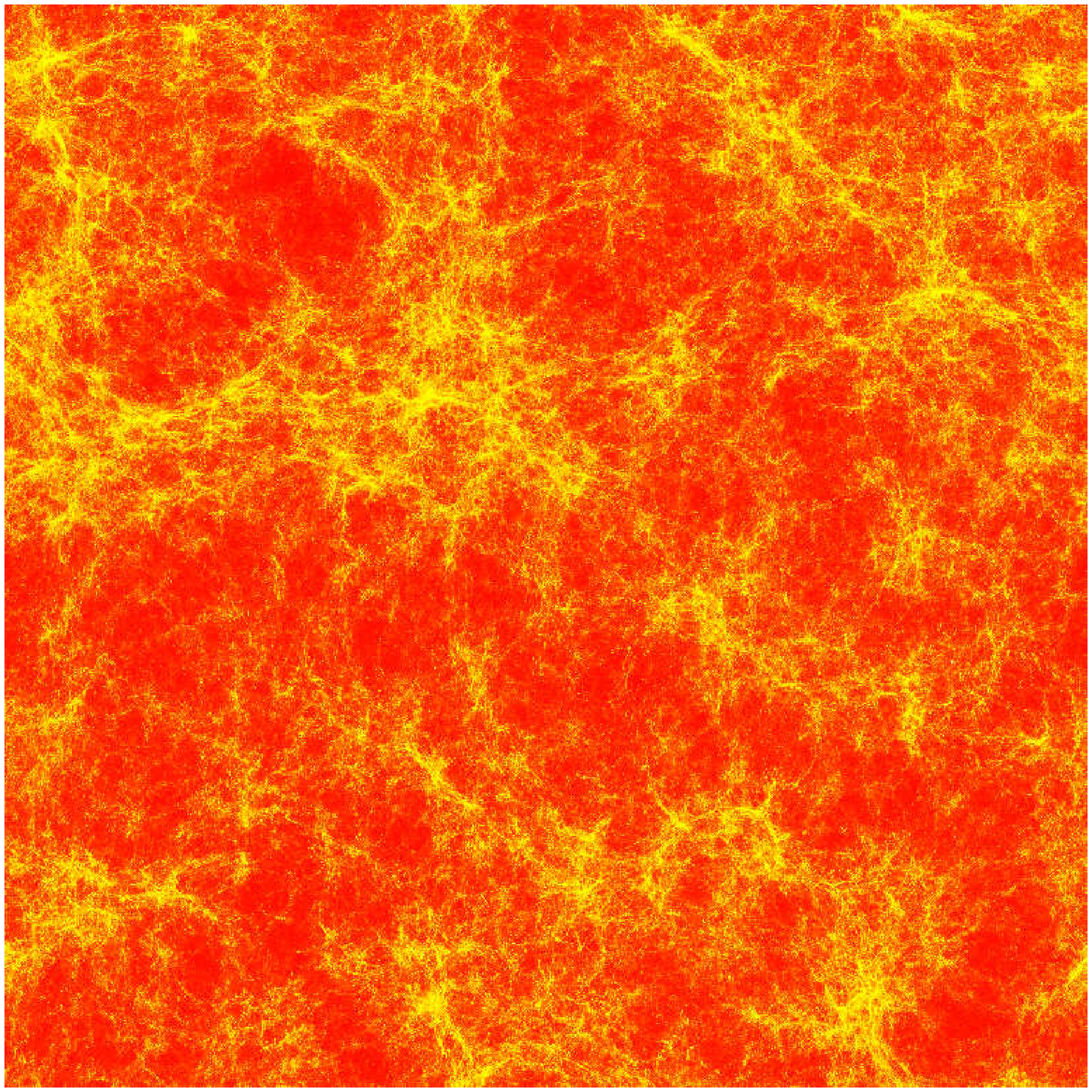}
    \includegraphics[width=3.0in]{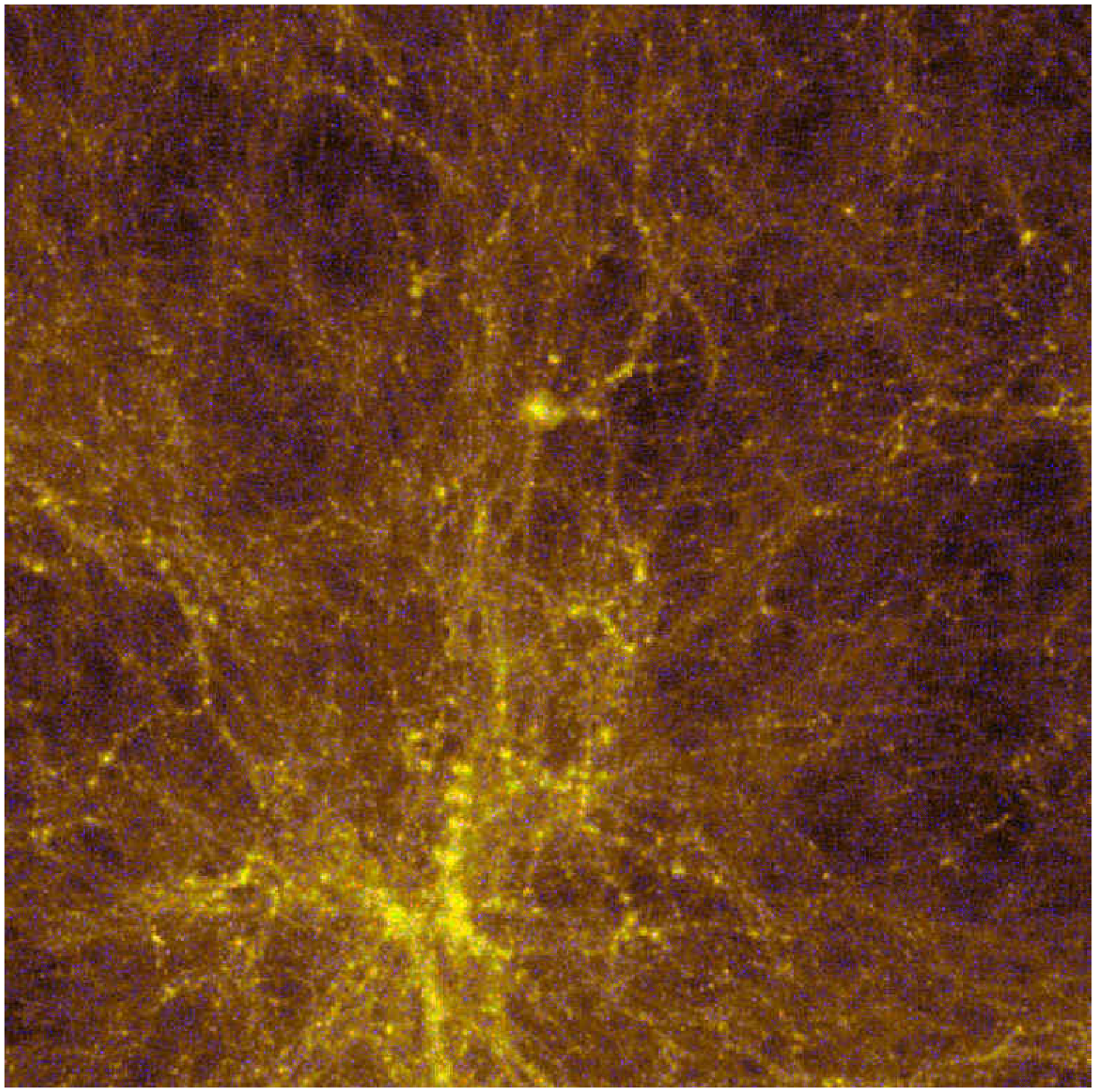}
  \end{center}
  \caption{(left) Slice of the Cosmic Web of Dark Matter at redshift $z=6$ from 
    our CubeP$^3$M Code simulation with $3072^3$ particles (29 billion) on a 
    $6144^3$ fine grid 
    in a comoving volume of 163~Mpc on a side.  
    (right) zoom-in of 10~Mpc $\times$ 10~Mpc subregion.
    Slices are 20 Mpc thick.
    \label{web}}
\end{figure*}

Reionization is generally believed to be the outcome of the release of
ionizing radiation by galaxies undergoing star formation (see for example
\citep{2006Sci...313..931B,2005SSRv..116..625C} for recent reviews). Current
theory suggests that the galaxies responsible for most of this
radiation are dwarf galaxies more massive than about $10^8M_\odot$. In
the Cold Dark Matter (CDM) universe nonlinear structures form from
initially small-amplitude, Gaussian-random-noise density fluctuations,
by a continuous hierarchical sequence of mergers and infall with the
smallest galaxies forming first and merging to yield larger ones,
forming the Cosmic Web of structures (Fig.~\ref{web}).  While dark
matter dominates the gravitational forces which cause this structure
formation, ordinary atomic matter must join the dark matter in making
galaxies for star formation to be possible.  Once the atomic gas in
the intergalactic medium (IGM) in some region is heated by
reionization, however, gas pressure opposes gravitational collapse,
and, thereafter, the smallest galaxies form without atomic matter and
cannot make stars. The minimum mass of star forming galaxies in such
regions is about one billion solar masses.

Due to the complex nature of the reionization process it is best
studied through large-scale numerical simulations
\citep{2006MNRAS.369.1625I, 2007ApJ...654...12Z, 2007ApJ...657...15K,
  2007ApJ...671....1T}. Such simulations present considerable
challenges related to the large dynamic range required and the
necessity to perform fast and accurate radiative transfer
calculations. The tiny galaxies which are the dominant contributors of
ionizing radiation must be resolved in very large cosmological
volumes, large enough to contain billions of times more total mass
than one dwarf galaxy and up to tens of millions of such galaxies, in
order to correctly derive their numbers and clustering properties. The
correct number densities and clustering of these ionizing sources
impact strongly the corresponding observational signatures. The
ionization fronts expanding from all these millions of galaxies into
the surrounding neutral IGM should then be tracked with a 3D radiative
transfer method and the full non-equilibrium chemistry should solved
in order to derive the resulting ionization state of the IGM. The
combination of all these requirements makes this problem a formidable
computational task.

We have recently performed the largest and most detailed simulations
 of the formation of early cosmological large-scale structures and
 their radiative feedback leading to cosmic reionization. This was
 achieved by running extremely large, $1728^3$- to $3072^3$-particle
 (5.2 to 29 billion) N-body simulations of the formation of
 the Cosmic Web, with enough particles and
 sufficient force resolution to resolve all the galactic halos with
 total masses larger than 100 million Solar masses in computational
 volumes of $(64/h\,\rm Mpc)^3$ to $(114/h\,\rm Mpc)^3$, respectively. 
 These results
 were then post-processed by propagating the ionizing radiation from all
 (up to millions of) sources by using fast and accurate ray-tracing
 radiative transfer method. For these simulations, we utilized a P$^3$M
 N-body code called CubeP$^3$M and a radiative transfer code called
 C$^2$-Ray, both developed by us. 

Our original simulations of the same process \citep{2006MNRAS.369.1625I} 
resolved the
formation of all galaxies more massive than about one billion solar
masses, those which are expected to form stars even after reionization
has heated their environment.  With this new generation of simulations
we explore the role of less massive dwarf galaxies. These are
important sources of ionizing radiation if they form before their
neighborhood is reionized but are prevented from being sources if they
form after it is reionized. Our preliminary results show that the
inclusion of these sources and of their suppression changes the
outcome of reionization substantially.


\section{CubeP$^3$M Code}

CubeP$^3$M is a massively-parallel P$^3$M (particle-particle-particle-mesh) 
\citep{Hockney:1988:CSU} cosmological N-body code written in Fortran 90. It 
is a successor of the code PMFAST \citep{2005NewA...10..393M} and shares 
many of its basic characteristics, particularly regarding the treatment of 
long-range (PM) gravitational forces. Here we briefly describe the 
code, focusing on the new elements which distinguish it from its 
predecessor and have not been previously discussed. The main new features 
are:
\begin{itemize}
\item Massive parallelization and scalability to thousands of nodes, using
(cubical) domain decomposition.
\item Inclusion of short-range direct particle-particle (PP) forces.
\end{itemize}
The code uses the FFTW 2.1.5 library and MPI-1, as well as (optional) 
OpenMP threading within each node in order to optimize the memory usage 
by reducing the amount of buffers necessary for the domain decomposition. 
CubeP$^3$M contains two mesh levels, a coarse and a fine mesh. The mesh sizes
per dimension are 2x the number of particles (fine mesh) and half the 
number of particles (coarse mesh). The superposition 
of the forces produced by each of these mesh levels, as well as the 
local particle-particle force, produces the total force on a particle in a 
given iteration of the code.


The computational domain is decomposed into equal-size cubical sub-regions, 
which are distributed across the nodes, with a 24-cell shared buffer zone 
around each domain. As a consequence of this decomposition CubeP$^3$M must 
be run on an integer-cubed number of MPI nodes. Each cuboid corresponds to 
the local section of the mesh, which is calculated in parallel by all nodes 
during each timestep. Since the coarse mesh is 4 times coarser per dimension 
than the fine mesh (which can be considered as the equivalent 1-level 
particle-mesh 
resolution), the total parallel workload is 64x less than the fine mesh 
simulation volume and does not suffer the performance bottleneck of having 
to calculate massive distributed FFTs. Each nodal cuboid is then decomposed 
into a number of fine mesh cuboid tiles for calculating the fine mesh forces. 
This is done to provide distinct work units for each of the threads on a 
node to process in parallel through multi-threading, as well as to decrease 
the amount of memory used to store the arrays associated with the fine mesh 
gravitational calculation.

We note that our approach of splitting the computational volume into
equal-size sub-domains, can in principle result in a work imbalance
between nodes and thus diminished efficiency. However, for
cosmological simulations of a sufficiently large volume of the
Universe this does not pose significant problems, as in practice the
variations of particle numbers between the (relatively large)
sub-volumes are fairly small.  The same would not be true for e.g.
detailed simulations of a single dark matter halo. We have thus geared
the code to the problem at hand, rather than trying to be universal.

Initial conditions are generated by a separate code using the standard
Zel'dovich approximation and primordial power spectrum transfer
function given by either the CMBFAST or CAMB Bolzmann code (both
publicly available). The main code is then started, with the following
algorithmic steps:
\begin{itemize}
\item Calculate the timestep for the current iteration, constraining
  it based on expansion (using the Friedmann equation) and the maximum
  acceleration components from the previous timestep.

\item Update the particle positions using the leapfrog method in
  half-timestep increments.  We actually merge the update from the
  second half of the previous step with the first half of the current
  step, as the velocities of the particles do not change

\item Construct a linked list for particles within the local volume
  based on their location within the coarse mesh.  This dramatically
  speeds up the particle searches which are used in the following steps.

\item Pass particles amongst nodes. Particles that move outside of
  the local physical volume are passed to adjacent nodes, as well as
  copies of local particles that are required to construct buffer
  densities.  Each dimension is treated independently and done
  synchronously, reducing the number of communication messages to 6
  from a possible 26 neighbors. As each pass is done the particles are
  added to the linked list.

\item Calculate fine mesh accelerations and apply to all particles in
  the same fashion as PMFAST \citep{2005NewA...10..393M}.

\item Calculate the particle-particle forces between particles within
  a fine-mesh grid cell and use the resulting acceleration to update
  their velocities.

\item Calculate the coarse mesh (long-range) accelerations and apply
  these to all particles in the same fashion as PMFAST.  As we use
  FFTW \citep{FFTW05}, which only includes a 1-dimensional (slab)
  decomposition MPI FFT implementation, we have written a wrapper
  layer that efficiently re-distributes the 3-dimensional (cubical)
  decomposition data to and from slab decomposition to perform the
  Fourier space convolution.

\item Update the particle positions at the end of the timestep and
  optionally checkpoint, calculate a density projection, or find
  halos. There are also options to output the coarse-mesh density,
  bulk velocity fields, and gas clumping for use in radiative-transfer
  simulations and to compare with observational predictions.

\item If we have reached the final redshift the simulation ends,
  otherwise we delete particles outside of physical volume and repeat
  these steps.

\end{itemize}

\begin{figure*}
  \begin{center}
    \includegraphics[width=3.3in]{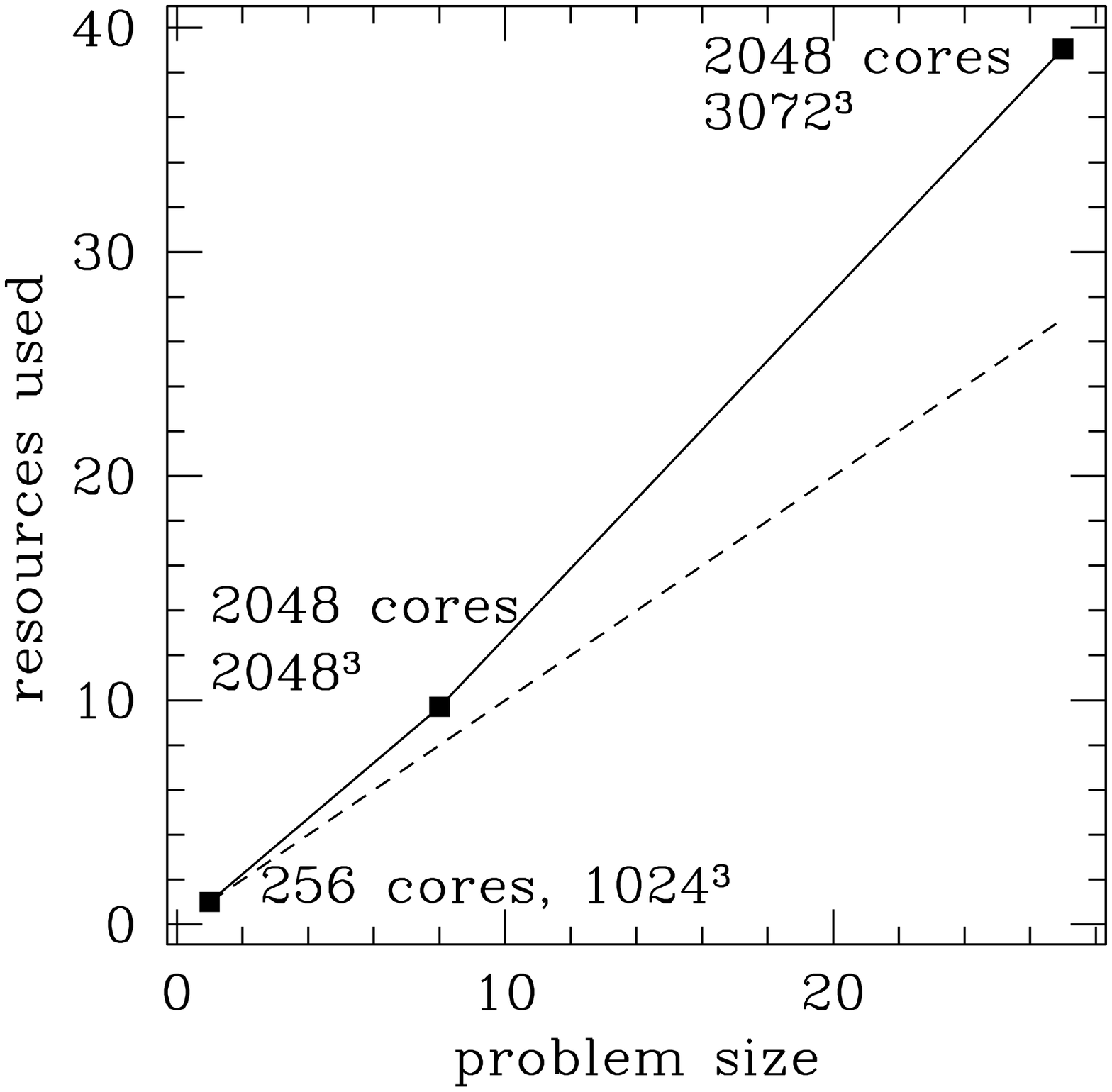}
    \includegraphics[width=3.3in]{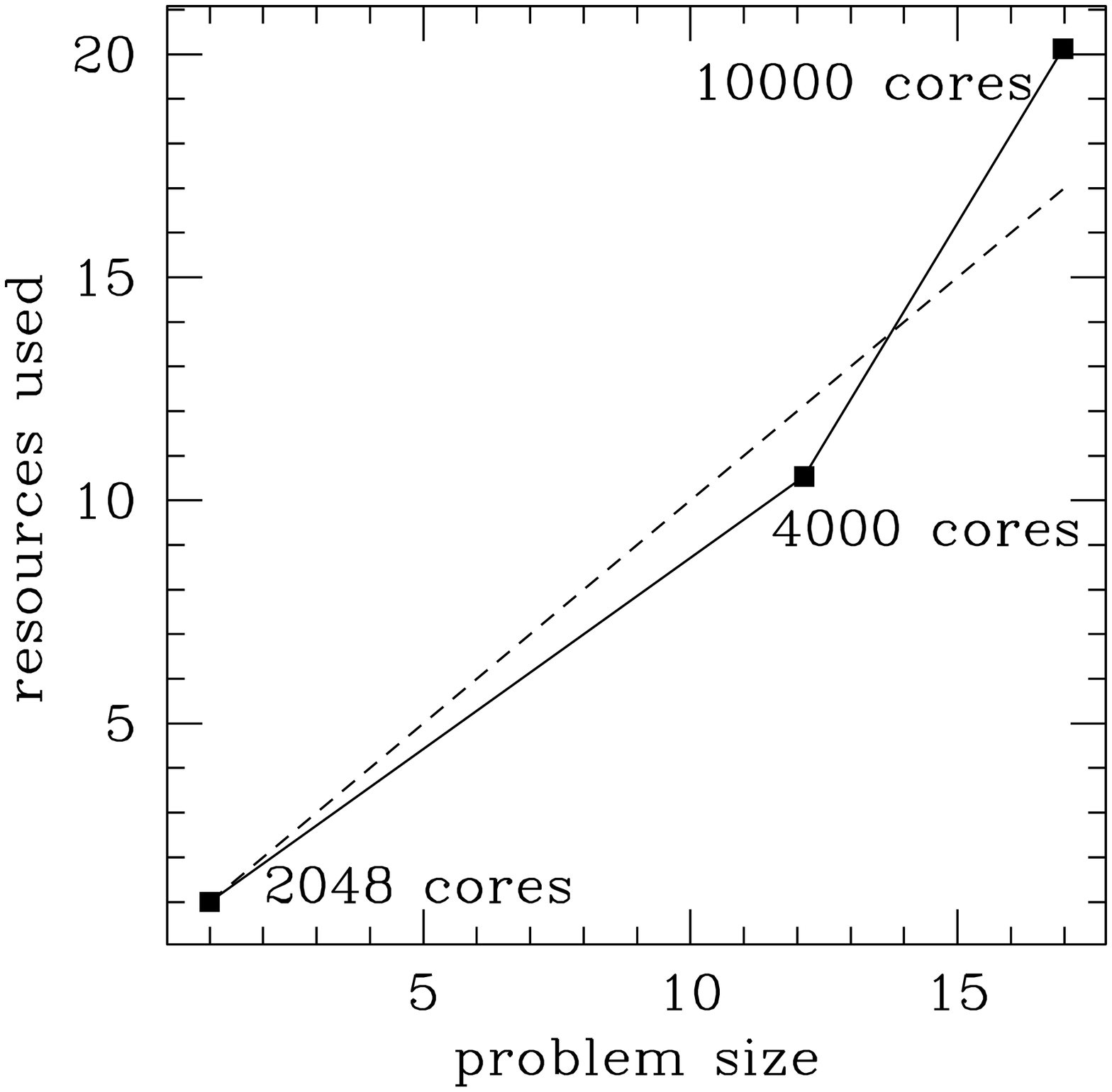}
  \end{center}
  \vskip -0.5cm 
  \caption{(a)(left) Scaling of the CubeP$^3$M code. Plotted are the
    computational resources used, (wall-clock time) $\times$(number
    of cores) vs. the problem size given by the total number of
    particles, both quantities normalized to the smallest of the
    three runs compared.  (b)(right) Scaling of the C$^2$-Ray code.
    Plotted are the computational resources used, (wall-clock
    time)$\times$(number of cores) vs. the problem size, (number of
    sources)$\times$(number of cells in grid), both quantities
    normalized to the smallest of the three runs compared.
    \label{scaling}}
\end{figure*}

The particle-particle force resolution at sub-grid distances is
achieved by including approximate direct particle-particle
interactions, as follows.  An NGP (nearest grid point) force kernel is
used to calculate the fine mesh force. This kernel has the property
that forces amongst particles within a grid cell are zero, and thus
provides the simplest method by which to match the particle-particle
forces with the fine mesh forces.

We calculate an additional PP force during fine mesh force velocity update, 
which allows us to avoid loading the particle list twice and also to thread 
the operation without any significant additional work. In this routine, all 
of the particles within a fine mesh tile are read in via the linked list and 
the appropriate force is applied depending on where they exist within the 
fine mesh tile. As the linked list chains particles within a coarse mesh cell, 
for the particle-particle interaction we proceed by constructing a set of 
fine-cell linked list chains for each coarse cell. We then proceed by 
calculating the particle-particle force between all particles within a fine 
mesh cell, limiting pairs to those exceeding a particular softening length 
(usually set to a twentieth of the initial inter-particle spacing) to prevent 
scattering. As this proceeds, we also accumulate the PP force applied to each 
particle and then determine the maximum force applied to a particular particle, 
which is then used to constrain the length of the global timestep.

The biggest problem with this method is that it is anisotropic and
depends on the location of the fine mesh with respect to the
particles. An example are two particles on either side of a grid cell
boundary. These particles will experience the 1-grid cell separation
NGP mesh force, but if the mesh were shifted such that they were
within the same cell, they would experience a larger force. We assume that 
on average this balances out, and we have even tuned the NGP force
kernel to provide as unbalanced of a force as possible at grid cell
distances by running large numbers of pairwise force comparisons at
different separations and positions relative to the mesh. This effect
is even more pronounced at the initial stages of the simulation where
the density is more homogeneous, and leads to mesh artifacts appearing
in the density field. In order to minimize this effect we have taken
to shifting the particle distribution relative to the mesh by a small
random amount, up to 2 fine grid cells in magnitude, in each
dimension, at each timestep. This adds negligible computational
overhead as it can be applied during the particle position update and
suppresses this behavior over multiple timesteps.

In the interest of speed and efficiency the halo catalogues are
constructed on-the-fly at a pre-determined list of redshifts. We use a
spherical over-density (SO) halo finder based on the density
distribution on the fine mesh \citep{1994MNRAS.271..676L}. The code
first builds the fine-mesh density for each tile. It then proceeds to
find and record all local density maxima within the physical volume
(excluding buffers) for the tile. It then uses parabolic interpolation
on the density field to determine the location of the maximum within
the cell, and records the peak position and value. Once the list of
maxima is generated they are sorted from highest to lowest. Then each
of the halo candidates is inspected independently starting with the
highest peak. The grid mass is accumulated in shells surrounding the
maximum until the average density of the halo drops below a
pre-defined overdensity cutoff (usually set to 178, in accordance with
the top-hat collapse model). As we accumulate mass we remove it from
the mesh, so that no mass is double-counted. This method is
inappropriate for finding sub-halos as those are naturally
incorporated in their host halos. After this stage we do another pass
through the halo list, this time using the mass measured above to
determine the radius of the halo and finding all particles which are
within this radius. Those are then used to calculate the halo bulk
velocity, internal velocity dispersion and angular momentum
components, all of which are listed in the final halo catalogues.

\section{C$^2$-Ray Code}

C$^2$-Ray (Conservative Causal Ray-Tracing) is a grid-based ray-tracing 
radiative transfer and non-equilibrium chemistry code. The method and our 
algorithm were described in detail in \citep{methodpaper}. The code is 
explicitly photon-conserving in both space and time, which ensures correct 
tracking of ionization fronts without loss of accuracy independent of the 
spatial and time resolution, with corresponding great gains in efficiency. 
The code has been tested in detail against a number 
of exact analytical solutions \citep{methodpaper}, as well as in direct 
comparison with a number of other independent radiative transfer methods on 
a standardized set of benchmark problems \citep{comparison1}. There is also 
a version of the code which is directly coupled to hydrodynamics evolution 
\citep{2006ApJ...647..397M}.

The ionizing radiation is ray-traced from every source to every grid cell 
using a short characteristics method, in which the neutral column density up 
to a given cell is given by interpolation of the column densities of the 
previous cells towards the source, in addition to the neutral column density 
through the cell itself. The contribution of each source to the local 
photoionization rate of a given cell is first calculated independently, 
after which all contributions are added together and then non-equilibrium 
chemistry solver is used to calculate the resulting ionization state. 
Normally multiple sources contribute to a local photoionization rate.  
Changes in the rate modify the neutral fraction and thus the neutral column 
density, which in turn changes the photoionization rates themselves (since 
either more or less radiation reaches the cell). An iteration procedure is 
thus called for in order to converge to the correct, self-consistent solution.

The previous version of the code has been used for simulations of
cosmic reionization and its observability
\citep[e.g.][]{2006MNRAS.369.1625I,
  2006MNRAS.372..679M,2007MNRAS.376..534I,kSZ,2007arXiv0711.2944I} and radiative
feedback in turbulent molecular clouds \citep{2006ApJ...647..397M}.
While our basic methodology remains essentially as described in
\citep{methodpaper}, recently the C$^2$-Ray has been thoroughly
re-written in Fortran 90, made more flexible and modular and
parallelized for distributed-memory machines. In terms of
parallelization strategy, due to the causal nature of the ray-tracing
procedure (i.e. the state of each cell can be calculated only after
all previous cells towards the source are done) it is not possible to
employ domain decomposition (except for a limited one, into octants,
see below), although other approaches exist which seek ways to
overcome this limitation \citep[]{2001MNRAS.321..593N,2006A&A...452..907R}.
Instead, the main code loop over the sources of ionizing radiation is
done in massively parallel fashion. Each MPI node has a copy of the
density field and receives a number of sources whose radiation to
trace through the grid.  The sources can be distributed over the nodes
either statically, or dynamically, using a master-slave approach,
whereby one node distributes the work amongst the rest. The static
allocation is preferable for relatively small number of nodes (up to
$\sim$32), while the dynamical approach is better for larger systems.
As long as the total number of sources is significantly larger than
the number of MPI nodes such parallelization is very efficient. For
the large-scale cosmological reionization problem there are typically
hundreds of thousands to millions of sources, thus our code scales
well up to tens of thousands of cores (see next section). For problems
with a (relatively) low number of ionizing sources such a
parallelization strategy would be inefficient, but such problems are
not sufficiently computationally-intensive to require massive
parallelization and could, instead, be solved on a smaller number of
nodes, or even in serial.  A similar situation occurs for the initial
steps of the simulations presented below, when the cosmological
structure formation is not yet much advanced, thus only a few to tens
of halos form. However, their number increases exponentially with
time, quickly reaching thousands, and then tens and hundreds of
thousands. We therefore start our simulations on a small number of
cores (typically 32), and increase this number as more sources form.
Since the current version of the code requires that a complete copy of
the grid is available on each node, it has relatively high memory-per-node
requirements. E.g. a $432^3$ mesh requires $\sim6.6$~GB per node to run.
Therefore, Ranger, with its large amounts of RAM available per node is 
a perfect platform for high-resolution radiative transfer runs.

As mentioned above, a limited domain decomposition into octants is possible 
for our method, since those are independent of each other within the 
short-characteristic ray-tracing framework. We use this to (optionally) 
improve the memory efficiency of the code by doing the grid octants 
in parallel within each MPI node using OpenMP multi-threading. This 
way each MPI node needs only one copy of the grid, which is shared 
amongst the cores within the node.

\section{Simulations and Scaling}

We have conducted a series of N-body structure formation simulations,
with particle numbers of $1024^3$, $1500^3$, $1728^3$, $2048^3$ and
$3072^3$ and box sizes of 53~Mpc, 80~Mpc, 91~Mpc, 103~Mpc and 163~Mpc,
respectively\footnote{These box sizes denote the so-called comoving size,
i.e.\ the physical size this volume will have at redshift 0 (the
current time).}. At earlier times the physical volume will be smaller by
a factor $1/(1+z)$. The last in this series is the largest simulation
of the formation of early cosmological structures done to date. 
The
$1500^3$ and $1728^3$-particle simulations were ran on the Dell Linux
Cluster (Lonestar) at TACC on 500 and 432 cores, respectively, while
the $1024^3, 2048^3$ and $3072^3$ were ran on the newly-deployed Sun
Constellation Linux Cluster (Ranger) at TACC, on 256, 2048 and 2048
cores, respectively. The latter three simulations were ran with 4 MPI
processes and 4 threads per node (using the {\it tacc\_affinity}
script to ensure local memory affinity) and required 4,100, 40,000 and
159,000 SUs (cores $\times$ hours).

All simulations use the same particle mass, $5\times10^6M_\odot$.
This mass resolution allows us to resolve, with 20 particles or
better, all halos with a mass above $10^8M_\odot$. This mass roughly
corresponds to a halo virial temperature of $10^4$~K, and thus is the
minimum halo mass for the gas in the halo to be able to cool through
hydrogen atomic-line cooling and efficiently form stars. Gas in halos
smaller than that can only cool through molecular, as well as
metal-line cooling, which is less efficient since molecules get
destroyed relatively easily, while metals (elements heavier than He)
are not present in significant amounts so early in the history of the
Universe.  The halos are divided further into halos with mass above
the Jeans mass of the ionized and heated medium (roughly above
$10^9M\odot$), which are not affected by reionization and lower-mass
halos, whose star formation is suppressed in ionized regions.
\begin{figure}
  \begin{center}
    \includegraphics[width=3.3in]{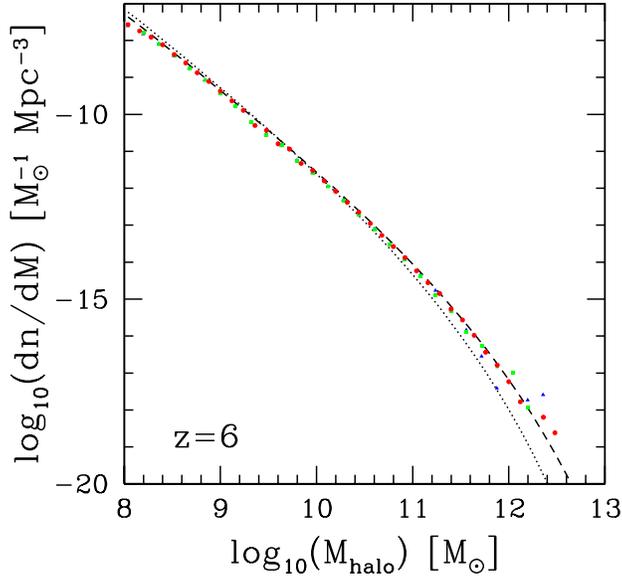}
  \end{center}
  \vskip -0.5cm
  \caption{Mass functions at $z=6$ from the N-body simulations with
    box sizes 91~Mpc (blue triangles), 103~Mpc (green
    squares) and 163~Mpc (red circles). All three mass
    functions demonstrate excellent agreement, except for the largest
    halos which are subject to cosmic variance for the smaller volume
    simulations. The lines show the predicted mass functions based on
    the well-known analytical models of Press-Schechter
    (\citep{1974ApJ...187..425P}; dotted) and Sheth-Tormen
    (\citep{2002MNRAS.329...61S}; dashed).
    \label{massf}}
\end{figure}

Based on the results from the N-body simulations, i.e.\ the halo catalogues and density field (re-gridded 
to a coarser grid), we run our radiative transfer reionization simulations. Our
methodology for these simulations is described in detail in  
\citep{2006MNRAS.369.1625I,2006MNRAS.372..679M,2007MNRAS.376..534I}. In short, all halos
are assumed to host galaxies (thus, ionizing sources), to which we assign 
ionizing photon luminosity proportional to the halo mass, with an efficiency
$f_\gamma$, giving the number of photons emitted per atom over the lifetime of
the source (here assumed to be 11.5 Myr). The simulations are denoted 
f100\_250S\_91Mpc\_216, f100\_250S\_91Mpc\_432 and f50\_250S\_163Mpc\_384, where
the first two numbers are the values of $f_\gamma$ for the high- and low-mass
sources (``S'' is for source suppression due to Jeans-mass filtering, the 
third is the box size and the last is the radiative transfer grid
resolution per dimension. The first two simulations are currently finished,
while the last (and largest) is still running. All three simulations ran on 
Ranger with 4 MPI processes and 4 threads per node and took approximately 
90,000, 700,000 and (to date) 500,000 SUs, respectively. 
\begin{figure}
  \begin{center}
    \includegraphics[width=3.6in]{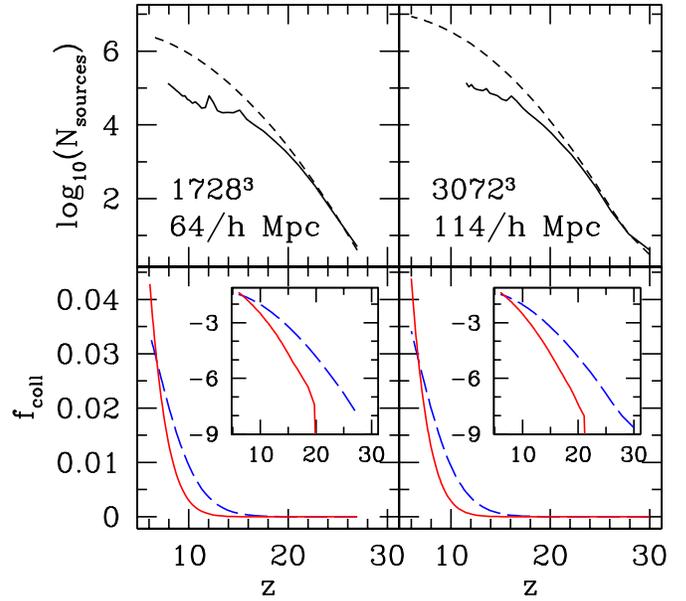}
  \end{center}
  \vskip -1cm
  \caption{(top panels) Total number of sources (dashed) and
    number of active sources (i.e. all sources less the suppressed
    ones; solid); and (bottom panels) collapsed fraction in
    high-mass (red, solid) and low-mass sources (blue, dashed)
    (insets show the same in log scale) vs. cosmic redshift for
    simulations f100\_250S\_91Mpc (right) and f50\_250S\_163Mpc (right).
    \label{sourcenum_fcoll}}
\end{figure}

Both CubeP$^3$M and C$^2$-Ray are designed for weak scaling, i.e.\ if
the number of cores and the problem size increase proportionally to
each other, for ideal scaling the wall-clock time should remain the
same, in contrast to ``strong'' scaling, whereby the same problem
solved on more cores should take proportionally less wall-clock time.
This weak scaling requirement is dictated by the problems we are
trying to solve (very large and computationally-intensive) and our
aims, which are to do such large problems efficiently, rather than for
the least wall-clock time. Some scaling results are shown in
Fig.~\ref{scaling}.

\begin{figure*}
  \begin{center}
    \includegraphics[width=3.3in]{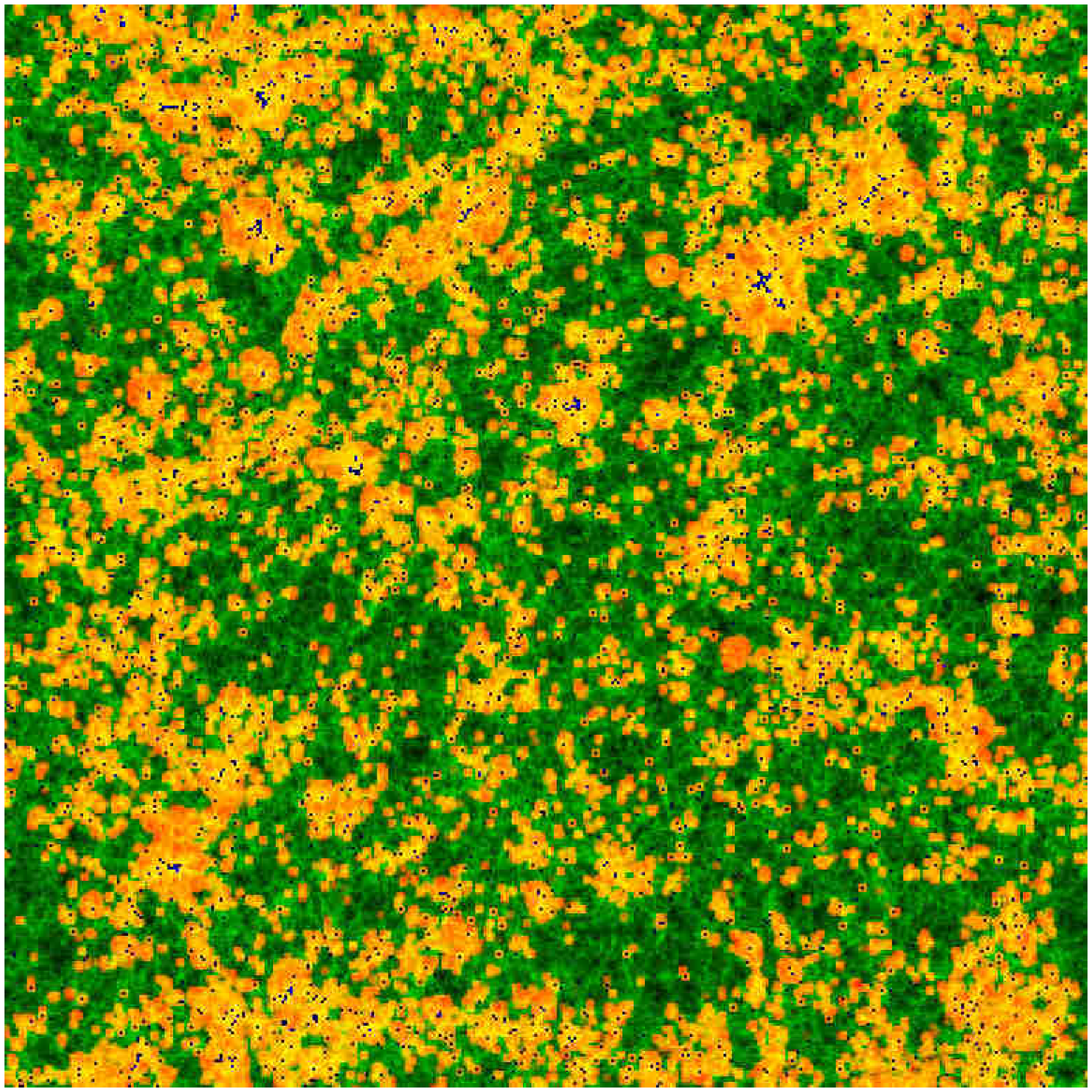}
    \hskip 1.5cm
    \includegraphics[width=3.3in]{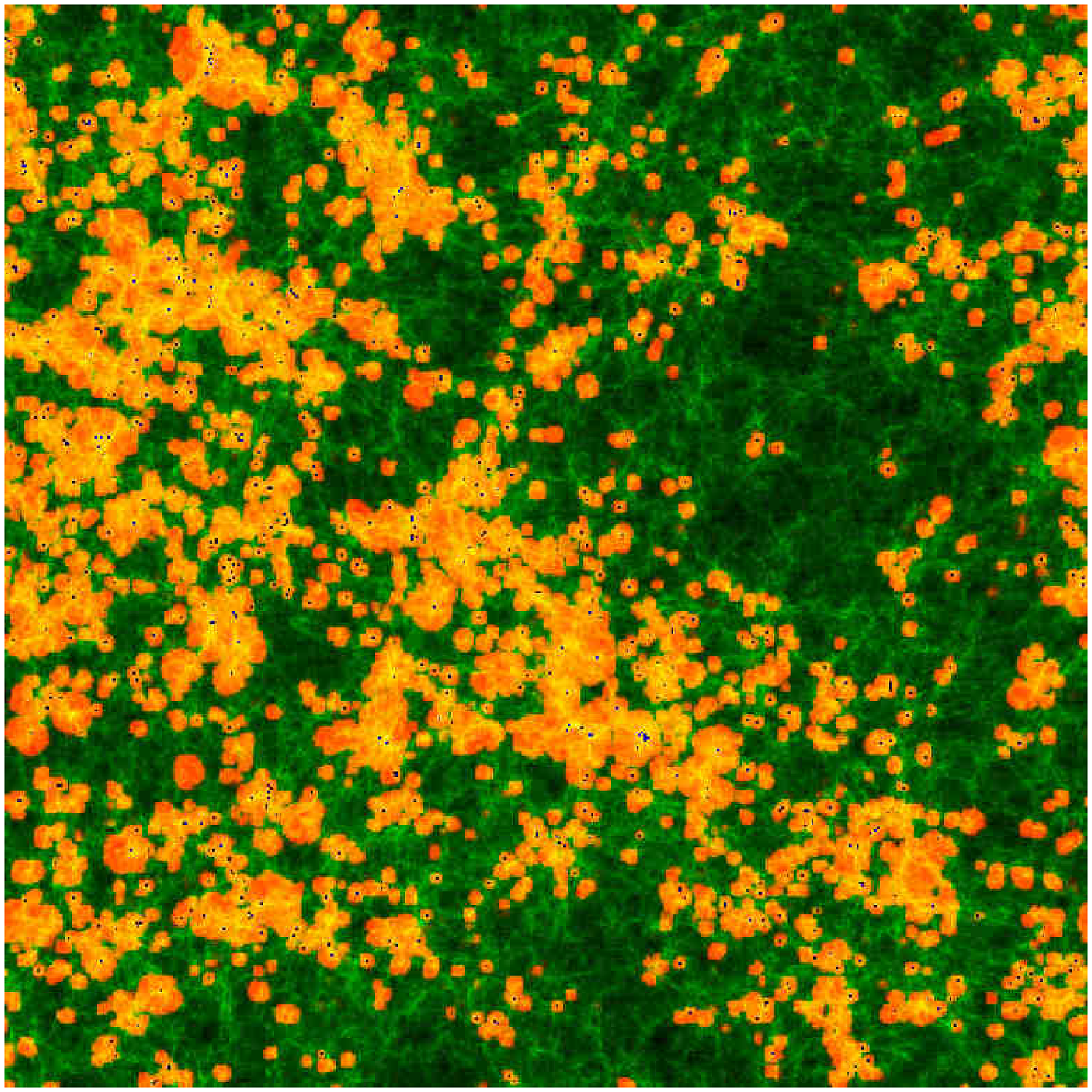}
    \vspace{-0.3cm}
  \end{center}
  \caption{Spatial slices of the ionized and neutral gas density from
    our radiative transfer simulations (a)(left) boxsize 
    163~Mpc at redshift $z=11.6$ and (b)(right) boxsize 
    91~Mpc  at $z=11.9$, both at ionized fraction by 
    mass $x_m=0.30$. Shown are the density field (green) overlayed with   
    the ionized fraction (red/orange/yellow) and the cells containing 
    sources (dark/blue). There is a good agreement between the two 
    simulations in the typical sizes of the locally-percolated ionization 
    bubbles. The slice thickness is 1/h Mpc for the 64/h Mpc box and 
    0.3/h Mpc (1 cell) for the 114/h Mpc box. The sources are projected 
    from thicker 
    slices: 1/h Mpc for the 64/h Mpc box and 3.3/h Mpc for the 114/h Mpc box.
    \label{images}}
\end{figure*}

For the N-body simulations we concentrate on the three Ranger runs, so
as to eliminate the complicating factor of comparing different
machines, clock speeds, etc. On the horizontal axis we show the size
of the problem (defined here as total number of particles, since all
three simulations have the same resolution), scaled to the smallest
one, while the vertical axis shows the resources used (SUs) for the
complete simulation (evolution from redshift $z=300$ to $z=6$).  The
code scaling from 256 to 2048 cores and $1024^3$ to $2048^3$ particles
is excellent, within 20\% of the perfect one (dashed line). We expect
that this small difference is mostly due to the larger communication
overhead for the larger simulation. The largest simulation ($3072^3$
particles on 2048 cores) differs from the perfect scaling by
approximately 40\%, demonstrating the strong effect of the
communication and I/O overheads (since the number of cores for the
$2048^3$ and $3072^3$ simulations is the same, the only difference is
the larger amount of data that has to be communicated and
read/written).  The code performance is thus quite satisfactory and it
could be expected to scale well to even larger number of cores.

The radiative transfer problem size scales proportionally to both the
grid size and the number of sources.
In this case we compare the code performance over a single timestep,
at $z\sim12$.  There are 40,797, 61,840, and 123,257 ionizing sources
for the cases f100\_250S\_91Mpc\_216, f100\_250S\_91Mpc\_432 and
f50\_250S\_163Mpc\_384, respectively. Results show almost perfect
scaling, within $\sim15\%$ from the ideal one, for up to 10,000 cores.
In fact, the scaling from 2048 to 4000 cores is even slightly better
than the ``perfect'' one, which reflects the fact that the fixed
component of the code, is a larger fraction of the total work for a
smaller grid (here $216^3$ vs. $432^3$). On the other hand, our
largest run, on 10,000 cores, scales slightly worse, partly due to the
same grid effect, but in reverse (the grid is $384^3$ vs. $432^3$),
and partly due to the higher communication overhead in this case.

\section{Results and Future Work}

The simulations presented here are still being analysed, but some
preliminary scientific results are shown in
Figs.~\ref{massf}-\ref{images}.  In Fig.~\ref{massf} we show the halo
mass functions at $z=6$ for three of our N-body simulations. All three
mass functions are in excellent agreement with each other, except for
some (expected) scatter for the largest, rarest halos, which are
subject to cosmic variance, particularly for the smaller boxes. The
numerical results are in good agreement with the
analytically-predicted Sheth-Tormen mass function
\citep{2002MNRAS.329...61S}, but not with the Press-Schechter one
\citep{1974ApJ...187..425P}, which strongly under-predicts the
abundances of the rare halos and over-predicts the abundances of the
low-mass ones (also see \citep{2006MNRAS.369.1625I,
  2007MNRAS.374....2R, 2007ApJ...671.1160L}). In
Fig.~\ref{sourcenum_fcoll} we show the evolution of the total source
numbers, active source numbers and collapsed fractions (i.e. the
fraction of the total mass found in sources) for the two types of
sources. The collapsed fractions from the two simulations agree to
within a few percent (relative) and the halo numbers show a similar
agreement (once the different simulation volumes are taken into
account). The biggest difference is found at very early times, when
only a few, rare sources have formed, which are subject to strong
cosmic variance. E.g.\ for the 163~Mpc box the first halos form
already at redshift $z\sim30$, while for 91~Mpc the first sources only
form at $z\sim27$, due to the 5.6 times smaller volume which is
simulated and, more subtly, due to the inevitable cutoff at the box
size of longest-wavelength density perturbation modes, which changes
the statistics and bias of the rarest halos.
  

Finally, in Fig.~\ref{images} 
we show some images of the
density field and (active) source distribution overlayed with the ionized 
fraction derived from our radiative transfer simulations. These indicate the 
very complex and evolving geometry and topology of the network of ionized 
bubbles and neutral patches, which needs to be studied in detail in order to
make predictions for the many observational experiments aiming
to detect these remote epochs.


We have planned and are working on a number of improvements of the codes 
and the simulations presented here. A very important development will be
to directly couple the CubeP$^3$M and C$^2$-Ray codes, which will allow
us to run simulations of the feedback of radiation on the gas in the 
forming cosmic structures. Such feedback is particularly important at 
small spatial scales and can modify significantly the forming structures, 
with corresponding observational implications. There are some significant 
issues to overcome in terms of the efficient parallelization of such coupled
code, due to the very different approaches currently employed for their
parallelization, discussed above.

An improvement for CubeP$^3$M would be extending the accuracy of the
N-body PP force resolution to multi-grid cell distances. This would
dramatically improve the force resolution and further minimize the
force anisotropy by allowing the matching to take place at larger
separations where the magnitude of the gravitational force is weaker.
While this would slow down the calculation, it would help to improve
the code efficiency (by increasing the number of calculations while
keeping the same amount of communication).

We are also working on improving our halo finding procedure. The current 
grid-based scheme works quite well for finding and determining the properties
of the larger halos (containing 40-50 particles or more), but is somewhat 
less precise for the smallest halos (down to 20 particles). Our new halo
finding method will be independent of the grid and thus more reliable for
identifying the smallest halos.

Finally, we also plan to run much larger N-body simulations, with somewhat 
different scientific goals. The largest of our current simulations, with 
29 billion particles, only used $\sim3\%$ of the available cores and memory
on Ranger. It is entirely within the machine and code specifications to run 
simulations with up to 165 billion particles (on 10,976 cores) in the near 
future. Even simulations with over 300 billion particles are in principle
possible, provided some current technical issues are resolved.
Such extremely large structure formation simulations will be
of great value for galaxy survey science since it becomes possible to 
simulate the whole observable Universe ($\sim(3\,h^{-1}$Gpc)$^3$ volume) in a 
single simulation with sufficient resolution to identify the typical
galaxies seen in large surveys (with luminosity $L_*$, roughly similar to 
our Milky Way Galaxy, or less).

\begin{acknowledgements}
  This work was supported in part by Swiss National Science Foundation
  grant 200021-116696/1, the Swedish Research Council grant 60336701,
  as well as by NSF grant AST 0708176, NASA grants NNX07AH09G and
  NNG04G177G, and Chandra grant SAO TM8-9009X.
\end{acknowledgements}


\bibliographystyle{plain}


\end{document}